\documentclass[pra,showpacs,superscriptaddress,amsmath,amssymb]{revtex4}
\usepackage{graphicx}

\begin{document}

\textsf{J. Phys. A: Math. Theor. 40 (2007) 155-161}

\title{Simple scheme for implementing the Deutsch-Jozsa algorithm in thermal cavity}

\author{Wen-Xing Yang} \email{wenxingyang2@126.com}
\affiliation{State Key Laboratory of Magnetic Resonance and Atomic and Molecular Physics, Center for Cold Atom Physics, Wuhan Institute of Physics and
Mathematics, Chinese Academy of Sciences, Wuhan 430071 China} \affiliation{Graduate School, Chinese Academy of Sciences, Beijing 100080, China}
\author{Zhe-Xuan Gong} \email{gongzhexuan@gmail.com}
\affiliation{Department of Physics, Huazhong University of Science and Technology, Wuhan 430074, China}
\author{Wei-Bin Li}
\affiliation{State Key Laboratory of Magnetic Resonance and Atomic and Molecular Physics, Center for Cold Atom Physics, Wuhan Institute of Physics and
Mathematics, Chinese Academy of Sciences, Wuhan 430071 China} \affiliation{Graduate School, Chinese Academy of Sciences, Beijing 100080, China}
\author{Xiao-Xue Yang}
\affiliation{Department of Physics, Huazhong University of Science and Technology, Wuhan 430074, China}

\date{\today}

\begin{abstract}
We present a simple scheme to implement the Deutsch-Jozsa algorithm based on two-atom interaction in a thermal cavity. The photon-number-dependent parts in the
evolution operator are canceled with the strong resonant classical field added. As a result, our scheme is immune to thermal field, and does not require the
cavity to remain in the vacuum state throughout the procedure. Besides, large detuning between the atoms and the cavity is not necessary neither, leading to
potential speed up of quantum operation. Finally, we show by numerical simulation that the proposed scheme is equal to demonstrate the Deutsch-Jozsa algorithm
with high fidelity.
\end{abstract}


\maketitle

Ever since Feynman first pointed out the concept of a quantum computer in 1982 \cite{1}, quantum computation has undergone rapid progress. The new type of
computer can solve some problems much faster than its classical counterpart. For example, the well-known Deutsch-Jozsa problem, which is to identify whether a
binary-valued function $f(x)$ of $N$ bits variables is \emph{constant} for all values of $x$, or \emph{balanced} (equal to 1 for exactly half of all the
possible $x$, and 0 for the other half), can be solved by using a single query of $f(x)$ on a quantum computer \cite{2,3}, whereas classical computer needs up
to $(2^{N - 1} + 1)$ queries \cite{4}. Till now, the Deutsch-Jozsa algorithm has been widely studied \cite{5}-\cite{9}, with its efficiency experimentally
tested \cite{10,11}.

The Deutsch-Jozsa algorithm can be briefly described as follows: Assume for the simplest case that $f(x)$ has only a one-bit input ($x=0$ or $1$). The
algorithm can be performed in a system consist of one query qubit 1 and an auxiliary working qubit 2, which are initially prepared in the superposed state:
\begin{equation}\label{eq1}
| \psi \rangle_{1} = \frac{1}{2}(| 0 \rangle _1 + | 1 \rangle _1 )(| 0 \rangle _2 - | 1\rangle _2 ).
\end{equation}

Then an unitary operator $U_{fn}$ is used to calculate $f(x)$, which acts as $U_{fn}|x\rangle_1 | y \rangle_2 = | x \rangle_1 | {y \oplus f(x)} \rangle_2 $.
Here $x,y \in \{0,1\}$, and $ \oplus $ indicates addition modulo $2$. The implementation of $U_{fn}$ on $| \psi_1 \rangle$ yields
\begin{equation}\label{eq2}
| \psi \rangle_{2} = \frac{1}{2}[( - 1)^{f(0)}| 0 \rangle _A + ( - 1)^{f(1)}| 1 \rangle _A ](| 0 \rangle _B - | 1 \rangle _B ).
\end{equation}

There are actually four possible $U_{fn}$: $U_{f1}$ corresponds to f(0)=f(1)=0; $U_{f2}$ corresponds to f(0)=f(1)=1; $U_{f3}$ corresponds to f(0)=0 and f(1)=1;
$U_{f4}$ corresponds to f(0)=1 and f(1)=0; After performing a Hadamard transformation on the query qubit, the state of qubit 1 will be in $|0\rangle$ for
$U_{f1}$ and $U_{f2}$, while for $U_{f3}$ and $U_{f4}$ it becomes $|1\rangle$. Thus a measurement on the query qubit will tell us whether $f(x)$ is constant or
balanced.

Recently, Zheng proposed a scheme \cite{12} for implementing the Deutsch-Jozsa algorithm in cavity QED, in which two atoms, one as a preparing qubit and one as
the query bit, sequentially interact with the cavity mode, which serves as the auxiliary working qubit. The main drawback of this scheme, as well as several
other implementations of quantum algorithms using cavity QED \cite{13,14}, is that they are sensitive to cavity decay or thermal field, which makes practical
experiment difficult to be scalable. Although the decoherence time of the cavity can be prolonged by keeping the cavity excited merely in a virtual way
\cite{15}, one has to make sure that the cavity remains always in the vacuum state throughout the procedure, otherwise it it still sensitive to thermal field
\cite{16}.

We notice that by resorting to a strong classical field, such drawbacks can be overcome \cite{17} in that the photon-number-dependent parts in the evolution
operator are canceled with resonant classical field added, thus rending immunity to thermal field. Based on the state evolution presented in Ref. \cite{17}, we
propose an improved scheme for implementing Deutsch-Jozsa algorithm which outruns the previous scheme in Ref. \cite{12} in three important aspects: (1) Both
the preparing atom and the auxiliary atom-level used in Ref. \cite{12} are not necessary; instead we only use two two-level atoms, which might simplify the
experimental procedures effectively if scalability is concerned; (2) Cavity is not required to remain in the vacuum state all of the time, and insensitivity to
thermal field is still insured; (3) No large detuning between atoms and the cavity is necessary, potentially giving rise to better performance of the speed and
fidelity of the whole procedure.

To describe our scheme, let us first consider two identical two-level atoms simultaneously interacting with a single-mode cavity field and driven by a
classical field. The Hamiltonian (assuming $\hbar = 1)$ in the rotating-wave approximation reads \cite{17}-\cite{19}
\begin{equation}
\begin{array}{l}
\label{eq3} H = \frac{1}{2}\sum\limits_{j = 1}^2 {\omega _0 \sigma _{z,j} } + \omega _a a^\dag a \\\quad\quad + \sum\limits_{j = 1}^2 {[g(a^\dag \sigma _j^ - +
a\sigma _j^ + ) + \Omega (\sigma _j^ + e^{ - i\omega t} + \sigma _j^ - e^{i\omega t})]},
\end{array}
\end{equation}
where $\sigma _{z,j} = | e \rangle _j { }_j\langle e | - | g \rangle _j { }_j\langle g |$, $\sigma _j^ + = | e \rangle _j { }_j\langle g |$, $\sigma _j^ - = |
g \rangle _j { }_j\langle e |$, with $| e \rangle _j (| g \rangle _j )$ being the excited (ground) state of the $j$th atom. $\omega _0 $, $\omega _a $ and
$\omega$ are the frequencies for atomic transition, cavity mode, and classical field, respectively. $a^\dag $ and $a$ are the creation and annihilation
operators for the cavity mode. $g$ is the atom-cavity coupling strength and $\Omega $ is the Rabi frequency of the classical field. Assume that $\omega _0 =
\omega $. Then we can obtain the following interaction Hamiltonian in the interaction picture:

\begin{equation}
\label{eq4} H_I = \sum\limits_{j = 1}^2 {[\Omega (\sigma _j^ + + \sigma _j^ - ) + g(e^{ - i\delta t}a^\dag \sigma _j^ - + e^{i\delta t}a\sigma _j^ + )]} ,
\end{equation}
where $\delta = \omega _0 - \omega _a $. For the new atomic basis $| \pm \rangle _j = (| g \rangle _j \pm | e \rangle _j ) / \sqrt 2$, we can rewrite $H_{I} =
H_{e}+H_{0}$ with
\begin{equation}
\label{eq5} H_0 = \sum\limits_{j = 1}^2 {2\Omega S_{z,j} },
\end{equation}
\begin{equation}
\label{eq6} H_e = \sum\limits_{j = 1}^2 {g[e^{ - i\delta t}a^\dag (S_{z,j} + \frac{1}{2}S_j^ - - \frac{1}{2}S_j^ + ) + H.c.]}.
\end{equation}

Here $S_{z,j} = (| + \rangle _j { }_j\langle + | - | - \rangle _j { }_j\langle - |) / 2$, $S_j^ + = | + \rangle _j { }_j\langle - |$ and $S_j^ - = | - \rangle
_j { }_j\langle + |$. Assuming that $\Omega \gg \delta ,g$, we can neglect the fast oscillating terms. Then the effective Hamiltonian $H_{e}$ reduces to
\begin{equation}
\label{eq7} H_e = g(e^{ - i\delta t}a^\dag + e^{i\delta t}a)\sigma _x,
\end{equation}
where $\sigma _x = \frac{1}{2}\sum\limits_{j = 1}^2 {(\sigma _j^ + + \sigma _j^ - )}$. The evolution operator for Hamiltonian (Eq. \ref{eq7}) can be written as:
\begin{equation}
\label{eq8}U_e (t) = e^{ - iA(t)\sigma _x^2 }e^{ - iB(t)\sigma _x a}e^{ - iC(t)\sigma _x a^\dag },
\end{equation}
which was first proposed for trapped-ion system \cite{20}. By solving the Schr\"{o}dinger equation $i\frac{dU_e (t)}{dt} = H_I U_e (t)$, we can obtain $B(t) =
g(e^{i\delta t} - 1) / i\delta$, $C(t) = - g(e^{ - i\delta t} - 1) / i\delta$ and $A(t) = g^2[t + (e^{ - i\delta t} - 1)/i\delta] / \delta$. Setting $\delta t
= 2\pi$, we have $B(t)=C(t)=0$. Then we can get the evolution operator of the system
\begin{equation}
\label{eq9}U_I(t) = e^{ - iH_0 t}U_e (t) = e^{ - i2\Omega \sigma _x t - i2\lambda \sigma _x^2 t},
\end{equation}
with $\lambda = g^2 / 2\delta $. We note that the evolution operator is independent of the cavity field state, allowing it to be in a thermal state. Unlike
Ref. \cite{15}, our scheme does not require $\delta\gg g$. The atoms interact with the cavity mode for a time $t$, leading to
\begin{equation}
\begin{array}{l}
\label{eq10} | e \rangle _1 | g \rangle _2 \to e^{ - i\lambda t}\{\cos (\lambda t)[\cos (\Omega t)| e \rangle _1 - i\sin (\Omega t)| g \rangle _1
]\\\\\quad\quad\quad\quad\times[\cos (\Omega t)| g \rangle _2 - i\sin (\Omega t)| e \rangle _2 ] - i\sin (\lambda t)\\\\\quad\quad\quad\quad\times[\cos (\Omega
t)| g \rangle _1 - i\sin (\Omega t)| e \rangle _1 ][\cos (\Omega t)| e \rangle _2
\\\\\quad\quad\quad\quad- i\sin (\Omega t)| g
\rangle _2 ]\}.
\end{array}
\end{equation}
and
\begin{equation}
\begin{array}{l}
\label{eq11} | g \rangle _1 | g \rangle _2 \to e^{ - i\lambda t}\{\cos (\lambda t)[\cos (\Omega t)| g \rangle _1 - i\sin (\Omega t)| e \rangle _1
]\\\\\quad\quad\quad\quad\times[\cos (\Omega t)| g \rangle _2 - i\sin (\Omega t)| e \rangle _2 ] - i\sin (\lambda t)\\\\\quad\quad\quad\quad\times[\cos (\Omega
t)| e \rangle _1 - i\sin (\Omega t)| g \rangle _1 ][\cos (\Omega t)| e \rangle _2
\\\\\quad\quad\quad\quad- i\sin (\Omega t)| g
\rangle _2 ]\}.
\end{array}
\end{equation}

We choose the interaction time and Rabi frequency appropriately so that $\Omega t =(2m+1) \pi $ ($m$ is an integer) and $\lambda t = \pi / 4$, then we obtain the
Einstein-Podolsky-Rosen state (EPR state)
\begin{equation}
\label{eq12} \frac{1}{\sqrt 2 }(| e \rangle _1 | g \rangle _2 - i| g \rangle _1 | e \rangle _2 ),
\end{equation}
and
\begin{equation}
\label{eq13} \frac{1}{\sqrt 2 }(| g \rangle _1 | g \rangle _2 - i| e \rangle _1 | e \rangle _2 ),
\end{equation}
where we have discarded the common phase factor. These calculations are useful in the later numerical analysis of fidelity.

Now we'd like to show that with the help of $U_I(t)$ as shown in Eq. \ref{eq9}, all the relevant unitary operators $U_{fn}$ in the Deutsch-Jozsa algorithm can
be easily achieved. But before that, we first need to show how to use the above idea to realize the quantum CNOT gate. For the present problem we let first atom
serve as the query qubit and the second atom as the auxiliary working qubit, \emph{i.e.} $|g\rangle_i$ and $|e\rangle_i$ respectively represent $|0\rangle_i$ and
$|1\rangle_i$ ($i=1,2$) in Eq. \ref{eq1}.

According to the Eq. \ref{eq9}, it can be easily shown that
\begin{equation}
\label{eq14} \left \{
\begin{array}{l}
 U_I (t)| + \rangle _1 | + \rangle _2 = e^{ -i2(\Omega + \lambda )t}| + \rangle _1 | + \rangle _2 \\\\
 U_I (t)| + \rangle _1 | - \rangle _2 = | +\rangle _1 | - \rangle _2 \\\\
 U_I (t)| - \rangle _1 | + \rangle _2 = | -\rangle _1 | + \rangle _2 \\\\
 U_I (t)| - \rangle _1 | - \rangle _2 = e^{ -i2(\Omega - \lambda )t}| - \rangle _1 | - \rangle _2
\end{array} \right.
\end{equation}

By setting $\delta = \sqrt 2 g$ and $gt = 2\pi $, we can make the interacting time $t$ and Rabi frequency $\Omega$ satisfy $\lambda t = \pi / 2$ and $\Omega t
= (2k + \frac{1}{2})\pi $ ($k$ is an integer). Then we have
\begin{equation}
\label{eq15} \left \{ {\begin{array}{l}
 U_I (t)| + \rangle _1 | + \rangle _2 = - | +\rangle _1 | + \rangle _2 \\\\
 U_I (t)| + \rangle _1 | - \rangle _2 = | +\rangle _1 | - \rangle _2 \\\\
 U_I (t)| - \rangle _1 | + \rangle _2 = | -\rangle _1 | + \rangle _2 \\\\
 U_I (t)| - \rangle _1 | - \rangle _2 = | -\rangle _1 | - \rangle _2
\end{array}} \right.
\end{equation}

As a result, we obtain a controlled-phase gate, which can be transformed into CNOT gate through a few single-qubit unitary operations:
\begin{equation}
\label{eq16}
\begin{array}{l}
 \quad\quad\quad \mbox{ }X_1 \quad\quad\quad \mbox{ }H_1 \quad\quad\quad \mbox{ }U_I (t)
 \quad\quad\quad \mbox{ }H_1 \quad\quad\quad\quad \mbox{ }X_1 \quad\quad\quad\quad \mbox{ }Z_1 \\
 | g \rangle _1 | g \rangle _2 \to | e\rangle _1 | g \rangle _2 \to | - \rangle _1 | g \rangle _2 \to \mbox{ }| - \rangle _1 | g \rangle _2 \to \mbox{ }| e
 \rangle _1 | g \rangle _2 \to \mbox{ }| g\rangle _1 | g \rangle _2 \mbox{ } \to | g \rangle _1 | g\rangle _2 \\\\
 | g \rangle _1 | e \rangle _2 \to | e\rangle _1 | e \rangle _2 \to | - \rangle _1 | e \rangle _2 \mbox{ } \to \mbox{ }| - \rangle _1 | e \rangle _2 \to \mbox{
 }| e \rangle _1 | e \rangle _2 \mbox{} \to \mbox{ }| g \rangle _1 | e \rangle _2 \mbox{ } \to \mbox{ }| g \rangle _1 | e\rangle _2 \\\\
 | e \rangle _1 | g \rangle _2 \to | g\rangle _1 | g \rangle _2 \to | + \rangle _1 | g \rangle _2 \to - | + \rangle _1 | e \rangle _2 \to - | g \rangle _1 | e
 \rangle _2 \to - | e \rangle _1 | e\rangle _2 \to | e\rangle _1 | e \rangle _2 \\\\
 | e \rangle _1 | e \rangle _2 \to | g\rangle _1 | e \rangle _2 \to | + \rangle _1 | e \rangle _2 \mbox{ } \to - | + \rangle _1 | g \rangle _2 \to - | g
 \rangle _1 | g \rangle _2 \to - | e \rangle_1 | g \rangle _2 \to | e\rangle _1 | g \rangle _2. \\\\
 \end{array}
\end{equation}

Here $H_1 $, $X_1 $, and $Z_1 $ are Hadamard, $\sigma _x $ and $\sigma _z $operations on the first atom with computational basis being $| g \rangle _1 $ and $|
e \rangle _1 $, which can be easily realized by choosing the appropriate amplitudes and phases of classical fields, respectively.

$U_{f1}$ operation: This operation on the atomic qubits does not require any interaction with the cavity mode. In this case the atoms can be tuned far off
resonant with the cavity mode and thus the atom-cavity evolution is freezing. Thus the system remains in the state of $|\psi\rangle_1$.

$U_{f2}$ operation: We first apply the aforementioned CNOT gate, and then perform the single-qubit transformation $| g \rangle _1 \to | e \rangle _1 $ and $| e
\rangle _1 \to - | g \rangle _1 $ on the atom $A$ by using a $\pi $-Ramsey pulse. Then we repeat the controlled-NOT operation and perform the transformation $|
g \rangle _1 \to - | e \rangle _1 $ and $| e \rangle _1 \to | g \rangle _1 $ by using a $\pi $-Ramsey pulse with a phase difference $\pi $ relative to the
first Ramsey pulse. Therefore we obtain
\begin{equation}
\begin{array}{l}
\label{eq17} | \psi \rangle_{2} = \frac{1}{2}(| 0 \rangle _1 + | 1 \rangle _1 )(| {0 \oplus 1} \rangle _2 - | {1 \oplus 1} \rangle _2 )\\\\ \quad\quad\ =
\frac{1}{2}( - | 0\rangle _1 - | 1 \rangle _1 )(| 0 \rangle _2 - | 1 \rangle _2 ).
\end{array}
\end{equation}

$U_{f3}$ operation: By performing the CNOT gate operation of Eq. \ref{eq16}, the output state of two-atom is given by
\begin{equation}
\begin{array}{l}
\label{eq18} | \psi \rangle_{2} = \frac{1}{2}[| 0 \rangle _1 (| 0 \rangle _2 - | 1 \rangle _2 ) + | 1 \rangle _1 (| {0 \oplus 1} \rangle _2 - | {1 \oplus 1}
\rangle _2 )]
\\\\\quad\quad\ = \frac{1}{2}(| 0 \rangle _1 - | 1
\rangle _1 )(| 0 \rangle _2 - | 1 \rangle _2 ).
\end{array}
\end{equation}

$U_{f4}$ operation: Firstly, we perform the single-qubit transformation $| g \rangle _1 \to | e \rangle _1 $ and $| e \rangle _1 \to - | g \rangle _1 $ on the
atom $A$; Secondly, the CNOT operation of Eq. \ref{eq16} is applied; Finally we perform the single-qubit transformation $| g \rangle _1 \to - | e \rangle _1 $
and $| e \rangle _1 \to | g \rangle _1 $. This leads to
\begin{equation}
\begin{array}{l}
\label{eq19} | \psi \rangle_{2} = \frac{1}{2}[| 1 \rangle _1 (| 0 \rangle _2 - | 1 \rangle _2 ) + | 0 \rangle _1 (| {0 \oplus 1} \rangle _2 - | {1 \oplus 1}
\rangle _2 )]
\\\\\quad\quad\ = \frac{1}{2}( - | 0 \rangle _1 + | 1
\rangle _1 )(| 0 \rangle _2 - | 1 \rangle _2 ).
\end{array}
\end{equation}

\begin{figure}
\includegraphics[width=9cm]{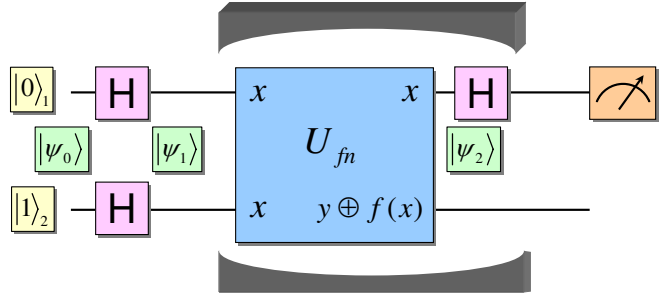}
\caption{\label{fig1} Experimental apparatus of the whole scheme, where atoms $A$ and $B$ cross the cavity with same velocity but at different positions, allowing for single-qubit operation of
each one in the process.}
\end{figure}

The whole scheme of the implementation of Deutsch-Jozsa algorithm is displayed in Fig. \ref{fig1}. Two atoms 1 and 2, first simultaneously prepared in box into
high lying circular Rydberg state denoted by $| g \rangle _1 | e \rangle _2 $, are in the initial average state after a Hadamard operation. Then they undergo
the operations in Fig. \ref{fig1} from left to right. In order to realize the different operations $U_{fn}$, we have to employ an inhomogeneous field to distinguish
the two atoms by the same trick as in Ref. \cite{13}. Finally the atoms 1 and 2 are separately read out by the state-selective field-ionization detectors.

We briefly discuss the experimental feasibility of our proposal. Although the evolution operator is independent of the thermal photons of cavity field as
decided by the condition $\delta t=2\pi$, the two-atom system is entangled with the cavity during the atom-cavity interaction. We have to neglect the cavity
decay during this interaction time. We assume that the atom-cavity coupling constant is $g = 2\pi \times 25$ kHz \cite{21,22}, $\delta = \sqrt {2}g$. Direct
calculation shows that the interaction time is at the order of $10^{-5}$s. Note that the photon decay time is $T_{c}\simeq 10^{-3}$s, thus much longer than the
interaction time. After the interaction, the atoms are disentangled with the cavity, that is, the operation will not be affected by the cavity decay during the
interaction time. Besides, the radiative time for the Rydberg atoms is $T_r=3\times 10^{-2} $s and the implementation time needed to complete the whole
procedure in the cavity is much shorter than $T_r $ as the time for single-qubit transformation is negligible. Thus the proposed scheme is realizable with the
present cavity QED techniques. The most probable difficulty is to send two atoms simultaneously through the cavity, but other work have shown that even though
there exist time difference, the negative influence is almost negligible. \cite{15,18}.

In the obtaining of Eq. \ref{eq7}, we have discarded the fast oscillating terms, which induce Stark shifts on the states  $| + \rangle _j$ and $| - \rangle _j$.
Here we numerically simulate the dependence of fidelity considering the error introduced by the Stark shift for generations of EPR state as in Eq. \ref{eq12}
with different values of detuning, as shown in Fig. \ref{fig2}(a). (Note that we have set $\Omega=20\delta$). The result from the plot shows that even for
$\delta=\sqrt{2}g$, the fidelity is still larger than 97{\%}, from which we know that large detuning is not required in our scheme. Besides, if we consider the
fluctuation of the Rabi frequency $\Delta\Omega=0.01\Omega$, directly calculation shows that the fidelity for the generation of EPR state decreases by only 0.02.

To check the feasibility of our scheme more strictly, we show in Fig. \ref{fig2}(b) the estimated achievable fidelity of the produced EPR state, where we consider the
existence of fluctuations in the atom-cavity interaction that leads to imperfections of the quantum Rabi pulses \cite{23}. The fidelity is plotted for
various strengths of imperfections in the Rabi pulses, where we assume for simplicity that the initial cavity state is in $| 5 \rangle $ and each pulse suffers
the same imperfection. The result from the plot tells that even for 10{\%} pulse error, the fidelity is still larger than 80{\%}. (Note that in real
experiments this kind of imperfection can be controlled around 3{\%}).

Furthermore, we show in Fig. \ref{fig2}(c) that if we assume that the cavity is initially in a Fock state $| n \rangle $, the success probability for producing EPR state
slightly decreases with the increase of photon number. Even for $n=10$, the fidelity is still larger than 99.5{\%}, which means that the whole process
is almost independent of the cavity field state.

\begin{figure*}\label{fig2}
\includegraphics[width=5.9cm]{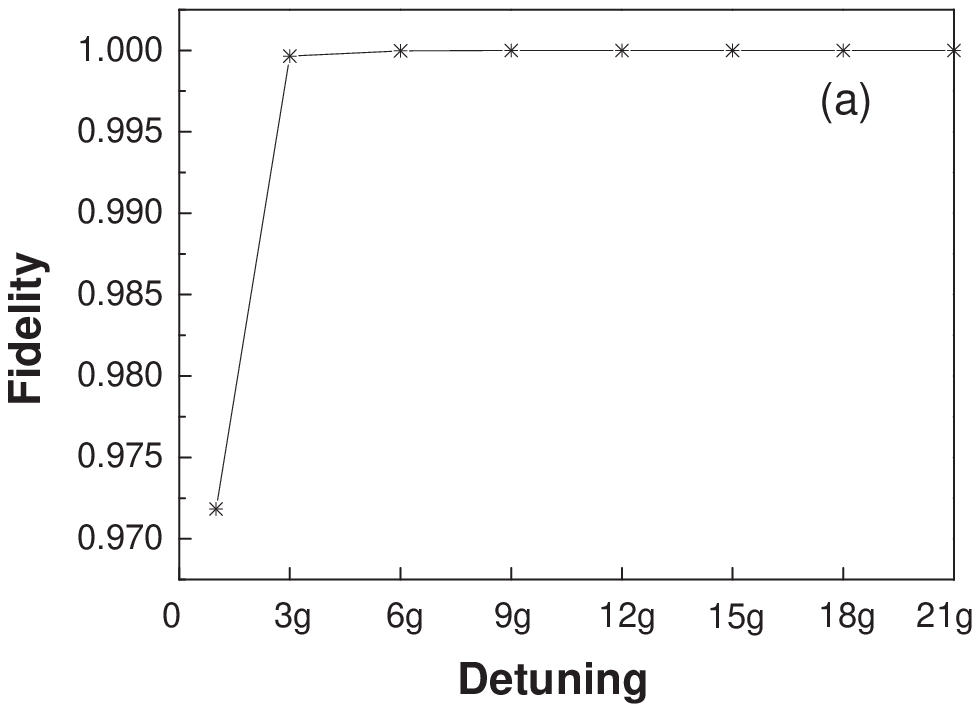}
\includegraphics[width=5.9cm]{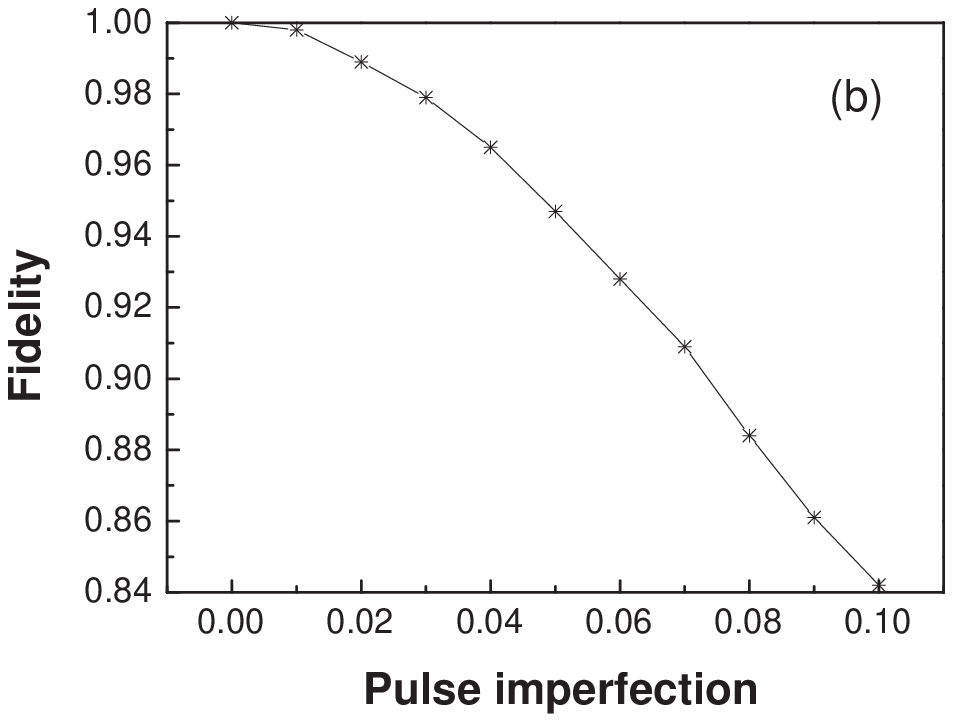}
\includegraphics[width=5.9cm]{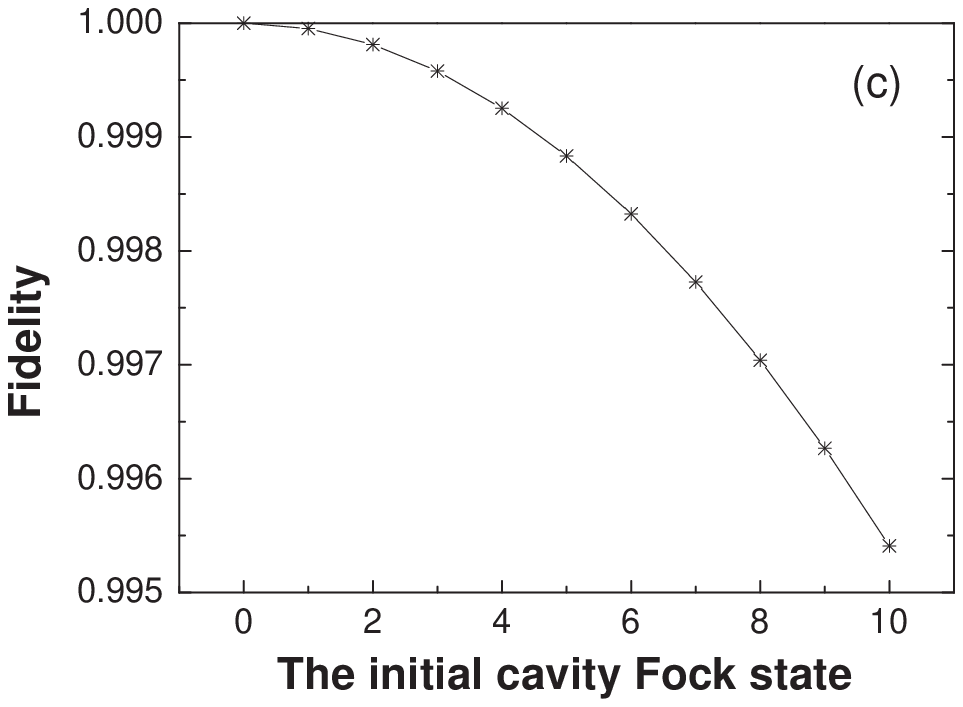}
\caption{\label{fig2} Numerical results for the fidelity of our scheme: (a) error introduced by the Stark Shifts ($\Omega=20\delta$); (b), (c) show the dependence of the
fidelity on pulse imperfections and initial cavity Fock state ($g=2\pi\times 25kHz$, $\delta=20\times g$, $\Omega=400\times g$).}
\end{figure*}

In principle, our scheme may offer a viable way to realize a scalable quantum algorithm. Based on the effective interaction between two atoms with single mode
cavity, our scheme can be also extended to multi-bit Deutsch-Jozsa algorithm, since the multi-bit entangled transformation $U'_{fn}$ can be constructed by single
qubit quantum gates and CNOT gates \cite{24}. However, it is still somewhat difficult for our scheme to be extended to many qubits based on current technology
\cite{22}. Furthermore, it should be pointed out that the single atom sources are required in our scheme.

To sum up, we have proposed a simple scheme for implementing the Deutsch-Jozsa algorithm in cavity QED based on effective interaction of two two-level atoms
with a single-mode cavity with the assistance of a strong classical driving field. Compared with the scheme in Ref. \cite{12}, our scheme is immune to thermal
field and does not require the cavity to remain in vacuum state. In addition, the scheme may work in a fast way since large detuning is not required. Besides,
our scheme does not require the auxiliary atom-level for the implementation of the quantum CNOT operation and all the operations except the single-qubit
transformation are imposed on both atoms simultaneously, making our scheme easier to be carried out practically. Based on these features, we present finally
the numerical analysis of the fidelity of our scheme with respect to the practical experiment under the influence of detuning, pulse imperfection and initial
cavity Fock state.

The authors acknowledgement Professor Ying Wu for many helpful discussions and encouragement. This work was partially supported by National Fundamental
Research Program of China 2005CB724508, by National Natural Science Foundation of China under Grant Nos. 60478029, 90503010, 10634060 and 10575040.

\end{document}